\documentclass[12pt]{article}
\pdfoutput=1
\usepackage{jheppub}
\usepackage{graphicx,epsfig,amsmath,amssymb}
\usepackage{subcaption}
\usepackage{enumitem}
\usepackage{calc}

\graphicspath{{./plots}}

\def\nn{\nonumber}

\def\be{\begin{equation}}
\def\ee{\end{equation}}
\def\bea{\begin{eqnarray}}
\def\eea{\end{eqnarray}}

\newcommand{\gosam}{\textsc{GoSam}{}}
\newcommand\POWHEG{{\tt POWHEG}}
\newcommand\POWHEGBOX{{\tt POWHEG-BOX}}
\newcommand\POWHEGBOXV{{\tt POWHEG-BOX-V2}}
\newcommand\herwig{{\tt HERWIG-7.2}}
\newcommand\pythia{{\tt PYTHIA-8}}
\newcommand{\mhh}{m_{hh}}

\newcommand{\pthh}{p_{T}^{hh}}
\newcommand{\ct}{c_t}
\newcommand{\ctt}{c_{tt}}
\newcommand{\chhh}{c_{hhh}}
\newcommand{\cg}{c_{ggh}}
\newcommand{\cgg}{c_{gghh}}
\newcommand{\ftapprox}{FT$_{\mathrm{approx}}$}
\title{A non-linear EFT description of $gg\to HH$ at NLO interfaced to POWHEG}

\author[a]{Gudrun Heinrich,}
\author[b]{Stephen P.~Jones,}
\author[c]{Matthias Kerner,}
\author[d]{Ludovic Scyboz}

\affiliation[a]{Institute for Theoretical Physics, Karlsruhe Institute of Technology (KIT), 76128 Karlsruhe, Germany}
\affiliation[b]{Theoretical Physics Department, CERN, Geneva, Switzerland}
\affiliation[c]{Physik-Institut, Universit{\"a}t Z{\"u}rich, Winterthurerstrasse 190, 8057 Z{\"u}rich, Switzerland}
\affiliation[d]{Rudolf Peierls Centre for Theoretical Physics, Parks Road, Oxford OX1 3PU, UK}

\emailAdd{gudrun.heinrich@kit.edu}
\emailAdd{s.jones@cern.ch}
\emailAdd{mkerner@physik.uzh.ch}
\emailAdd{ludovic.scyboz@physics.ox.ac.uk}

\preprint{{\small  CERN-TH-2020-106, KA-TP-04-2020, OUTP-20-06P, P3H-20-026, ZU-TH-21/20}}

\abstract{
We present the implementation of Higgs boson pair production in gluon
fusion within a non-linear Effective Field Theory framework containing five
anomalous couplings for this process. The code, available within the
\POWHEGBOXV{},  includes full NLO QCD corrections with massive top quarks.
All five couplings can be modified by the user.
We show $\mhh$ distributions at seven benchmark points provided by an
$\mhh$ shape analysis at NLO and showered $\pthh$ distributions resulting from 
an interface to \pythia{} and \herwig.}
 
\keywords{Higgs couplings, Monte Carlo generators, EFT, NLO}

\begin{document}

\maketitle
\newpage


\section{Introduction}

Precise measurements of the Higgs boson couplings to other particles and itself are among the main goals for the next phases of LHC and beyond.
As the precision of the measurements increases, it is of great importance to have Standard Model predictions well under control, and to have reliable simulations of the effects of anomalous couplings.
In fact, measurements of Higgs couplings to electroweak bosons and the top quark are already reaching a level where systematic uncertainties play an increasingly important role~\cite{Aad:2019mbh,Sirunyan:2018sgc}. 
The trilinear Higgs-boson self-coupling $\chhh$ still is rather weakly constrained, however the window of possible $\chhh$-values has been narrowed considerably in Run II~\cite{Sirunyan:2018two,Aad:2019uzh}.

\medskip

Higgs boson pair production in gluon fusion in the SM has been calculated at leading order in Refs.~\cite{Eboli:1987dy,Glover:1987nx,Plehn:1996wb}. 
The NLO QCD corrections with full top quark mass dependence became available more recently~\cite{Borowka:2016ehy,Borowka:2016ypz,Baglio:2018lrj,Baglio:2020ini}.
The NLO results of Refs.~\cite{Borowka:2016ehy,Borowka:2016ypz} have been combined with parton shower Monte Carlo programs in Refs.~\cite{Heinrich:2017kxx,Jones:2017giv,Heinrich:2019bkc}, where Ref.~\cite{Heinrich:2019bkc} allows the trilinear Higgs coupling to be varied.

Before the full NLO QCD corrections became available, the $m_t\to\infty$ limit, sometimes also called ``Higgs Effective Field Theory~(HEFT)'' approximation or ``Heavy Top Limit (HTL)'',  has been used. 
In this limit, the NLO corrections were first calculated in 
Ref.~\cite{Dawson:1998py} using the so-called ``Born-improved HTL'', 
which involves rescaling the NLO results in the $m_t\to\infty$ limit by a factor $B_{\rm FT}/B_{\rm HTL}$, where $B_{\rm FT}$
denotes the LO matrix element squared in the full theory.
In Ref.~\cite{Maltoni:2014eza} an approximation called
``\ftapprox'', was introduced, which contains the real radiation matrix elements 
with full top quark mass dependence, while the virtual part is
calculated in the Born-improved $m_t\to\infty$ approximation.

In the $m_t\to\infty$ limit, the NNLO QCD corrections have been computed in Refs.~\cite{deFlorian:2013uza,deFlorian:2013jea,Grigo:2014jma,deFlorian:2016uhr}.
 The calculation of Ref.~\cite{deFlorian:2016uhr} has been combined with results including the top quark mass dependence as far as available in Ref.~\cite{Grazzini:2018bsd}, and soft gluon resummation on top of these results has been presented in Ref.~\cite{deFlorian:2018tah}. N$^3$LO corrections have become available recently~\cite{Chen:2019lzz,Chen:2019fhs}, where in Ref.~\cite{Chen:2019fhs} the N$^3$LO results in the heavy top limit have been ``NLO-improved'' using the results of Refs.~\cite{Heinrich:2017kxx,Heinrich:2019bkc}.

The scale uncertainties at NLO are still at the 10\% level, while they are decreased to about 5\% when including the NNLO corrections and to 
about 3\% at N$^3$LO in the ``NLO-improved'' variant.
The uncertainties due to the chosen top mass scheme have been assessed in Refs.~\cite{Baglio:2018lrj,Baglio:2020ini}.

For a more detailed description of the various developments and phenomenological studies concerning Higgs boson pair production we refer to recent review articles, e.g. Refs.~\cite{Amoroso:2020lgh,Cepeda:2019klc,DiMicco:2019ngk,Dawson:2018dcd}.



The main purpose of this paper is to present an update of the public Monte Carlo event generator {\tt POWHEG-BOX-V2/ggHH}, where the user can choose the values of  five anomalous couplings relevant to Higgs boson pair production as input parameters.
It is based on the implementation of the fixed-order NLO results~\cite{Borowka:2016ehy,Borowka:2016ypz}, combined with a non-linear Effective Field Theory framework~\cite{Buchalla:2018yce},  in the 
\POWHEGBOX~\cite{Nason:2004rx,Frixione:2007vw,Alioli:2010xd}. It builds on the code described in Ref.~\cite{Heinrich:2019bkc} which allows variations of the trilinear Higgs coupling (and the top Yukawa coupling) only.
We also show results for seven benchmark points, which have been identified by an $\mhh$ shape analysis presented in Ref.~\cite{Capozi:2019xsi}, based on the full NLO calculation, and compare NLO effects to effects from anomalous couplings.
Further, we show results matched to the \pythia~\cite{Sjostrand:2014zea} and \herwig~\cite{Bellm:2017bvx} parton showers to enable the assessment of parton-shower related uncertainties.

This paper is organised as follows. In Section~\ref{sec:couplings} we  describe the theoretical framework and the definition of the anomalous couplings. In Section~\ref{sec:usage} we describe the code and the usage of the program within the \POWHEGBOXV. Section~\ref{sec:results} contains the discussion of phenomenological results, before we conclude in Section~\ref{sec:conclusions}.
More detailed usage instructions are given in an appendix.

\section{Anomalous couplings in Higgs boson pair production}
\label{sec:couplings}

The calculation builds on the ones presented in Refs.~\cite{Buchalla:2018yce,Heinrich:2019bkc} and therefore will be described only briefly here. 

\begin{figure}[h]
\begin{center}
\includegraphics[width=11cm]{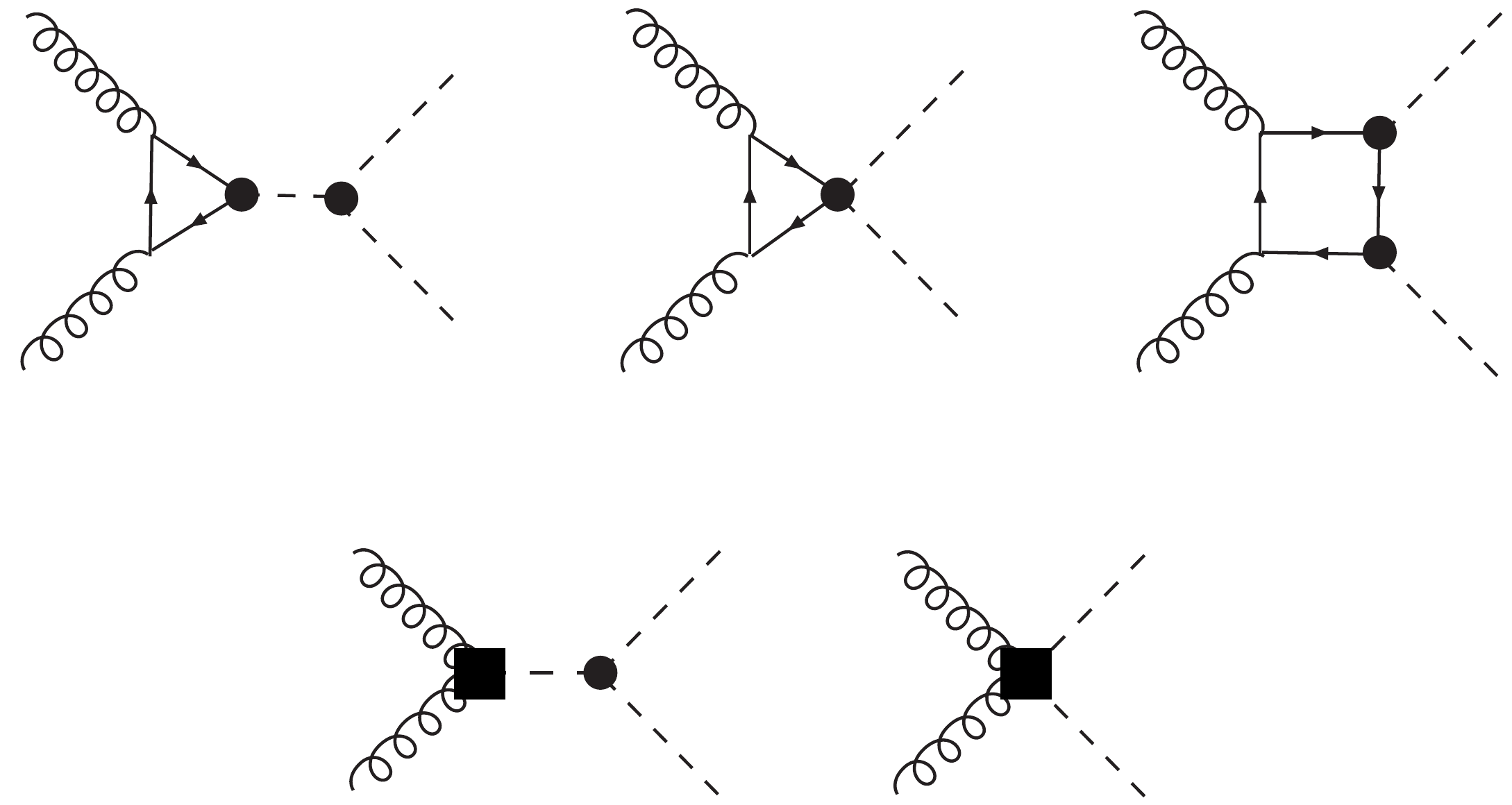}
\end{center}
\caption{Higgs boson pair production in gluon fusion at leading order
in the chiral Lagrangian. The black dots indicate vertices from
anomalous couplings present already at leading order in the Lagrangian,
the black squares denote effective interactions from contracted loops.}
\label{fig:hprocess}
\end{figure}

We work in a non-linear EFT framework, sometimes also called
Electroweak Chiral Lagrangian (EWChL)
including a light Higgs boson~\cite{Alonso:2012px,Buchalla:2013rka}.
It relies on counting the chiral dimension of the terms
contributing to the Lagrangian~\cite{Buchalla:2013eza}, rather than counting the canonical
dimension as in the Standard Model Effective Field Theory (SMEFT). In this way, the EWChL is also suitable for describing strong dynamics in the Higgs sector.
Applying this framework to Higgs boson pair production in gluon
fusion, keeping terms up to chiral dimension four, we obtain  the
effective Lagrangian relevant to this process as
\begin{align}
{\cal L}\supset 
-m_t\left(c_t\frac{h}{v}+c_{tt}\frac{h^2}{v^2}\right)\,\bar{t}\,t -
c_{hhh} \frac{m_h^2}{2v} h^3+\frac{\alpha_s}{8\pi} \left( c_{ggh} \frac{h}{v}+
c_{gghh}\frac{h^2}{v^2}  \right)\, G^a_{\mu \nu} G^{a,\mu \nu}\;.
\label{eq:ewchl}
\end{align}
In the EWChL framework there are a priori no relations between the
couplings. In general, all couplings may have arbitrary values of ${\cal O}(1)$.
The conventions are such that in the SM $c_t=c_{hhh}=1$ and $c_{tt}=c_{ggh}=c_{gghh}=0$.
The leading-order diagrams are shown in Fig.~\ref{fig:hprocess}.

In Ref.~\cite{Buchalla:2018yce}  the NLO QCD
corrections were calculated within this framework, and NLO results were presented for the
twelve benchmark points defined in Ref.~\cite{Carvalho:2015ttv}.

In Ref.~\cite{Capozi:2019xsi}, shapes of the Higgs
boson pair invariant mass distribution $\mhh$ were analysed in the 5-dimensional
space of anomalous couplings using machine learning techniques to
classify $\mhh$-shapes, starting from NLO predictions.
In more detail,  $10^5$ NLO distributions were produced to train a
neural network based on an autoencoder which extracts common shape
features, like an enhanced tail or a double peak, in the $\mhh$
distribution.
Then a {\tt KMeans} clustering algorithm from {\tt  scikit-learn}~\cite{scikit} was used to identify
distinct shape clusters. The aim was to produce clusters that distinguish characteristic shape features
without picking on minor details.
The cluster centres in the coupling parameter space resulting from
this procedure were then associated with candidate benchmark points.
However, if the corresponding total cross section exceeded the limit of
  $6.9\times \sigma_{SM}$~\cite{Aad:2019uzh}, which is currently the
  most stringent bound on the total cross section, we proceeded to the
  parameter point corresponding to the curve next-closest to the cluster center.
This method led to seven new benchmark points being identified, which we use here
to discuss our phenomenological results. For convenience we repeat the
benchmark points in Table~\ref{tab:benchmarks}, together with the
corresponding values for the cross section at  $\sqrt{s}=14$\,TeV.

\begin{table}
\begin{tabular}{|c|c|c|c|c|c|c|c|c|}
\hline
benchmark & $c_{t}$ & $c_{hhh}$ & $c_{tt}$ & $c_{ggh}$ & $c_{gghh}$ & $\sigma_{\rm{NLO}}$ [fb] & K-factor & ratio to SM \\
\hline
SM & 1 &  1 &  0 & 0 &  0 & 32.90 $\pm$ 0.03 & 1.66 & 1.00 \\
\hline
1 & 0.94 &  3.94 &  -$\frac{1}{3}$ & 0.5 &  $\frac{1}{3}$ & 222.63 $\pm$ 0.12 & 1.90 & 6.77 \\
\hline
2 & 0.61 & 6.84 & $\frac{1}{3}$ &  0.0 & -$\frac{1}{3}$ & 168.13 $\pm$ 0.07 & 2.14 & 5.11 \\
\hline
3 & 1.05 &  2.21 &  -$\frac{1}{3}$ & 0.5 & 0.5 & 151.94  $\pm$ 0.09 & 1.83 & 4.62 \\
\hline
4 & 0.61 &  2.79 &  $\frac{1}{3}$ &  -0.5 & $\frac{1}{6}$ &63.14 $\pm$ 0.03 & 2.15 & 1.92 \\
\hline
5 & 1.17 &  3.95 &  -$\frac{1}{3}$ & $\frac{1}{6}$ &  -0.5 & 154.77 $\pm$ 0.23 & 1.63 & 4.70 \\
\hline
6 & 0.83 &  5.68 &  $\frac{1}{3}$ &  -0.5 &  $\frac{1}{3}$ & 179.35 $\pm$ 0.18 & 2.16 & 5.45 \\
\hline
7 & 0.94 & -0.10 &  1& $\frac{1}{6}$ &  -$\frac{1}{6}$ &  131.06 $\pm$ 0.08 & 2.28 & 3.98 \\
\hline
\end{tabular}
\caption{NLO benchmark points derived in
  Ref.~\cite{Capozi:2019xsi}. The values for the cross section are given at
  $\sqrt{s}=14$\,TeV.}
\label{tab:benchmarks}
\end{table}

There are different normalisation conventions for the anomalous couplings in the literature.
In Table~\ref{tab:conventions} we summarise the conventions commonly used.
\begin{table}[htb]
\begin{center}
\begin{tabular}{ |c | c |c| }
\hline
Eq.~(\ref{eq:ewchl}), i.e. ${\cal L}$ of Ref.~\cite{Buchalla:2018yce}& Ref.~\cite{Carvalho:2015ttv}&Ref.~\cite{Grober:2015cwa}\\
\hline
$c_{hhh}$ & $\kappa_{\lambda}$ & $c_3$\\
\hline
$c_t$ &$ \kappa_t$  &$c_t$\\
\hline
$ c_{tt} $ & $ c_{2}$  &$c_{tt}/2$\\
\hline
$c_{ggh}$ &$ \frac{2}{3}c_g $  &$8c_{g}$\\
\hline
$c_{gghh}$ & $-\frac{1}{3}c_{2g}$ &$4c_{gg}$\\
\hline
\end{tabular}
\end{center}
\caption{Translation between the conventions for the definition of the anomalous couplings.\label{tab:conventions}}
\end{table}

We also give the relation to the corresponding parameters in the SMEFT, using the following Lagrangian based on the counting of canonical dimensions:
\begin{align}\label{eq:lsmeft}
\Delta{\cal L}_6 &=
\frac{\bar c_H}{2 v^2}\partial_\mu(\phi^\dagger\phi)\partial^\mu(\phi^\dagger\phi)
+\frac{\bar c_u}{v^2} y_t(\phi^\dagger\phi\, \bar q_L\tilde\phi t_R +{\rm h.c.})
-\frac{\bar c_6}{2 v^2}\frac{m^2_h}{v^2} (\phi^\dagger\phi)^3
\nonumber\\
&+\frac{\bar c_{ug}}{v^2} g_s
(\bar q_L\sigma^{\mu\nu}G_{\mu\nu}\tilde\phi t_R +{\rm h.c.})
+\frac{4\bar c_g}{v^2} g^2_s \phi^\dagger\phi\, G^a_{\mu\nu}G^{a\mu\nu}\;,
\end{align}
where we follow the conventions used in \cite{Giudice:2007fh,Grober:2015cwa},
except for $\bar c_g$ which differs by inclusion of the weak coupling $g^2$:
$\bar c_g\Big|_{\text{Ref.\cite{Grober:2015cwa}}}=g^2\,\bar c_g\Big|_{\text{Eq.}(\ref{eq:lsmeft})}$.
The term proportional to $\bar c_{ug}$ denotes the chromomagnetic operator,
which does not contribute at the order in the chiral counting we are considering here ($d\chi\leq 4$),
because it gets an additional loop suppression factor $1/16\pi^2$ due to the fact that dimension-6 operators involving field strength tensors (such as $\sigma^{\mu\nu}G_{\mu\nu}$) can only be generated through loop diagrams~\cite{Arzt:1994gp,Buchalla:2018yce}.
The remaining coefficients $\bar c_i$ in Eq.~(\ref{eq:lsmeft}) can be related to the couplings of the physical
Higgs field $h$ and compared with the corresponding parameters of the
chiral Lagrangian (\ref{eq:ewchl}).
After a field redefinition of $h$ to eliminate $\bar c_H$ from the
kinetic term one finds~\cite{Azatov:2015oxa,Grober:2015cwa}
\begin{align}
 c_t&=1-\frac{\bar c_H}{2}-\bar c_u\; ,\;
c_{tt}=-\frac{\bar c_H + 3\bar c_u}{4}\; ,\;c_{hhh}=1-\frac{3}{2}\bar c_H +\bar c_6 \;,\label{cthhh}\\
c_{ggh}&=2 c_{gghh}=(16\pi^2)\times 8\bar c_g\;. \label{cg}
\end{align}

\section{Description of the code}
\label{sec:usage}

\subsection{Structure of the code}

The code is an extension of the one presented in Ref.~\cite{Heinrich:2019bkc} to
include the possibility of varying all five anomalous couplings rather
than only the trilinear Higgs coupling.
For the virtual two-loop corrections,  we have built on the results of the
calculations presented in Refs.~\cite{Borowka:2016ehy,Borowka:2016ypz}.
These results were obtained by performing a (partial) reduction of the two-loop amplitude to master integrals based on {\sc Reduze}~\cite{Studerus:2009ye,vonManteuffel:2012np} and a subsequent numerical evaluation of the master integrals using the program {\sc SecDec}~\cite{Borowka:2015mxa,Borowka:2017idc}.
This has been done with the top quark- and Higgs masses fixed to a numerical value. Therefore these mass values should not be changed in the {\tt ggHH} code.

The real radiation matrix elements were implemented
using the interface between \gosam~\cite{Cullen:2011ac,Cullen:2014yla} and
the \POWHEGBOX~\cite{Alioli:2010xd,Luisoni:2013cuh}.
The extra matrix elements occurring in the EFT framework have been
generated by \gosam{} via a model file in {\tt
  UFO} format~\cite{Degrande:2011ua} which has
been developed in Ref.~\cite{Buchalla:2018yce}, 
derived from the effective Lagrangian in Eq.~(\ref{eq:ewchl}) using {\sc FeynRules}~\cite{Alloul:2013bka}.

The framework presented in Ref.~\cite{Heinrich:2019bkc} to interface the two-loop 
virtual contribution in {\tt POWHEG} is generalised in the following way: instead 
of a second-order polynomial (as for variations of $\chhh$ only), at NLO we can write the 
squared matrix element for variations of all five anomalous couplings as in Eq.~(\ref{eq:AcoeffsNLO}),
following Refs.~\cite{Azatov:2015oxa,Carvalho:2015ttv,Buchalla:2018yce}.

\begin{align}
|\mathcal{M}_{\rm BSM}|^2  &=   a_1\, c_t^4 + a_2 \, c_{tt}^2  + a_3\,  c_t^2 \chhh^2  +
a_4 \, \cg^2 \chhh^2  + a_5\,  \cgg^2  +
a_6\, c_{tt} c_t^2 + a_7\,  c_t^3 \chhh \nn\\
& + a_8\,  c_{tt} c_t\, \chhh  + a_9\, c_{tt} \cg \chhh + a_{10}\, c_{tt} \cgg +
a_{11}\,  c_t^2 \cg \chhh + a_{12}\, c_t^2 \cgg \nn\\
& + a_{13}\, c_t \chhh^2 \cg  + a_{14}\, c_t \chhh \cgg +
a_{15}\, \cg \chhh \cgg\, + a_{16}\, c^3_t \cg \nn\\
& + a_{17}\,  c_t c_{tt} \cg
+ a_{18}\, c_t \cg^2 \chhh + a_{19}\, c_t \cg \cgg \, + a_{20}\,  c_t^2 \cg^2
\nn\\
&+ a_{21}\, c_{tt} \cg^2
+ a_{22}\, \cg^3 \chhh + a_{23}\, \cg^2 \cgg  \,.
\label{eq:AcoeffsNLO}
\end{align}

For the Born-virtual interference term, we produce grids using 6715 points (5194 points at $\sqrt{s}=14$ TeV and 1521 points at $100$ TeV)
for 23 linearly independent sets of couplings. This enables us to 
derive, for each phase-space point, the coefficients $a_1, \dots, a_{23}$ by 
interpolation. Once the user has chosen a set of anomalous couplings, the 23 grids are
combined into one using Eq.~(\ref{eq:AcoeffsNLO}). This step is performed only once, in 
the first {\tt POWHEG} parallel stage. The Born and real contributions are evaluated exactly
for the chosen anomalous couplings without relying on a grid or interpolation. Note that
the $a_i$ coefficients of Eq.~(\ref{eq:AcoeffsNLO}) are not equal to the $A_i$ coefficients of 
Ref.~\cite{Buchalla:2018yce}, which are derived for the (normalised) cross-section and 
not for the Born-virtual interference term.

The original phase space points were produced with SM couplings,
therefore some regions, for example the low $\mhh$-region, are less
populated than the peak of the $\mhh$-distribution in the SM.
The statistical uncertainty on the input data induces a systematic uncertainty
on the 23 grids.
We have checked the relative size of the uncertainties of the virtual
corrections in each bin of the $\mhh$ distributions for all seven
benchmark points. This uncertainty is below 2\% throughout the
whole $\mhh$ range, except for the first bin.
This bin is poorly populated in the SM and therefore the uncertainties
in this bin are larger for coupling configurations where the low-$\mhh$
region is very different from the SM case or where the relative size of the virtual contribution is large. 
Hence we find uncertainties in the first $\mhh$ bin of about 6\%  for all benchmarks except 
for benchmark 5, which has a 12\% uncertainty in this bin.


\subsection{Usage of the code}

The code  can be found at the web page
\begin{center}
 \url{http://powhegbox.mib.infn.it}
\end{center}
under {\tt User-Processes-V2} in the {\tt ggHH} process directory.
An example input card ({\tt powheg.input-save}) and a run script ({\tt run.sh}) 
are provided in the {\tt testrun} folder accompanying the code.

In the following we only describe the input parameters that are specific to the
process $gg\to HH$ including five anomalous couplings. The
parameters that are common to all \POWHEGBOX{} processes can be found
in the \POWHEG{} manual {\tt V2-paper.pdf} in the \POWHEGBOXV{}/{\tt Docs} directory.

\subsubsection*{Running modes}
The code contains the SM NLO QCD amplitudes with full top quark mass dependence.
A detailed description of the different approximations can be found in
Ref.~\cite{Borowka:2016ypz}. For the Standard Model case as well as for BSM-values of the trilinear
Higgs coupling $\chhh$, the code can be run in four different modes, either by
changing the flag {\tt mtdep} in the \POWHEGBOX{} run card {\tt
  powheg.input-save}, or by using the script {\tt run.sh [mtdep
  mode]}.
If all five anomalous couplings are varied, there is only the
possibility of either calculating at

\begin{itemize}
\item NLO with full top quark mass dependence, or
\item LO (setting 
{\tt bornonly=1}) in either the full theory or in the $m_t\to\infty$ limit.
\end{itemize}

In more detail, the following choices are available:
\begin{description}
 \item[{\tt mtdep=0}:]{all amplitudes
   are computed in the $m_t\to\infty$ limit (HTL).} {\em This option is only available at NLO in the SM
   case or if only $\chhh$ is varied, or at LO.}
 \item[{\tt mtdep=1}:]{computation using Born-improved HTL. In this
   approximation the fixed-order part is computed at NLO in the heavy top limit
   and reweighted pointwise in the phase-space by the LO matrix
   element with full mass dependence divided by the LO matrix
   element in the HTL.} {\em This option is only available at NLO in the SM
   case or if only $\chhh$ is varied, or at LO.}
 \item[{\tt mtdep=2}:]{computation in the approximation \ftapprox. In
   this approximation the matrix elements for the Born and the real
   radiation contributions are computed with full top quark mass dependence, whereas the virtual part is
   computed as in the Born-improved HTL. }{\em This option is only available at NLO in the SM
   case or if only $\chhh$ is varied, or at LO.}
 \item[{\tt mtdep=3}:]{NLO computation with full top quark mass dependence.}
\end{description}

\subsubsection*{Input parameters}

The bottom quark is considered massless in all four {\tt mtdep} modes. The Higgs
bosons are generated on-shell with zero width. A decay can be attached
through the parton shower in the narrow-width approximation. However,
the decay is by default switched off (see the {\tt hdecaymode} flag in the
example {\tt powheg.input-save} input card in {\tt testrun}).

The masses of the Higgs boson and the top quark are set by default to
$m_h=125$\,GeV, $m_t=173$\,GeV, respectively, and their widths
have been set to zero. The full SM two-loop virtual contribution has
been computed with these mass values hardcoded, therefore they should not be changed when running with {\tt mtdep = 3}, otherwise the two-loop virtual part would contain a different top or Higgs mass from the rest of the calculation.
 It is no problem to change the values of $m_h$
and $m_t$ via the {\tt powheg.input-save} input card when running with
{\tt mtdep} set to $0$, $1$ or $2$.

The Higgs couplings as defined in the context of the Electroweak Chiral Lagrangian (see~\cite{Buchalla:2018yce} and references within), can be varied directly in the {\tt powheg.input} card. These are, with their SM values as default:
\begin{description}[leftmargin=!,labelwidth=\widthof{{\tt cgghh=0.0}:}]
\item[{\tt chhh=1.0}:] { the ratio of the Higgs trilinear coupling to its SM value,}
\item[{\tt ct=1.0}:] { the ratio of the Higgs Yukawa coupling to the top quark to its SM value,}
\item[{\tt ctt=0.0}:] { the effective coupling of two Higgs bosons to a top quark pair,}
\item[{\tt cggh=0.0}:] { the effective coupling of two gluons to the Higgs boson,}
\item[{\tt cgghh=0.0}:] { the effective coupling of two gluons to two Higgs bosons.}
\end{description}

 The possibility of varying all Higgs couplings (rather than $\chhh$ only)  is only
 available in the mode {\tt mtdep=3} (full NLO).
 More details about the  {\tt mtdep=3} running mode are given in
 Appendix \ref{sec:appendix1}.

 The runtimes are dominated by the evaluation of the real radiation
 part. When run in the full NLO mode, the runtimes we observed were in
 the ballpark of 100 CPU hrs for an uncertainty of about 0.1\% on the
 total cross section.

\section{Phenomenological results}
\label{sec:results}

We present results calculated at a centre-of-mass energy of
$\sqrt{s}=14$\,TeV using the
PDF4LHC15{\tt\_}nlo{\tt\_}30{\tt\_}pdfas~\cite{Butterworth:2015oua,CT14,MMHT14,NNPDF}
parton distribution functions interfaced to our code via
LHAPDF~\cite{Buckley:2014ana}, along with the corresponding value for
$\alpha_s$.  The masses of the Higgs boson and the top quark have been
fixed to $m_h=125$\,GeV, $m_t=173$\,GeV and their widths have been set to zero.
The top quark mass is renormalised in the on-shell scheme.
Jets are clustered with the
anti-$k_T$ algorithm~\cite{Cacciari:2008gp} as implemented in the
FastJet package~\cite{Cacciari:2005hq, Cacciari:2011ma}, with jet
radius $R=0.4$ and a minimum transverse momentum 
$p_{T,\mathrm{min}}^{\rm{jet}}=20$\,GeV.  The scale uncertainties are
estimated by varying the factorisation/renormalisation scales
$\mu_{F}, \mu_{R}$, where the bands 
represent 3-point scale variations around the central scale $\mu_0 =\mhh/2$, with
$\mu_{R} = \mu_{F}=c\,\mu_0$, where $c \in \{0.5,1,2\}$.
For the case $c_{hhh}=c_{hhh}^{\mathrm{SM}}=1$ we checked that the bands
obtained from these variations coincide with the bands resulting from
7-point scale variations.

\begin{figure}[htb]
  \centering
  \begin{subfigure}{0.49\textwidth}
    \includegraphics[width=\textwidth]{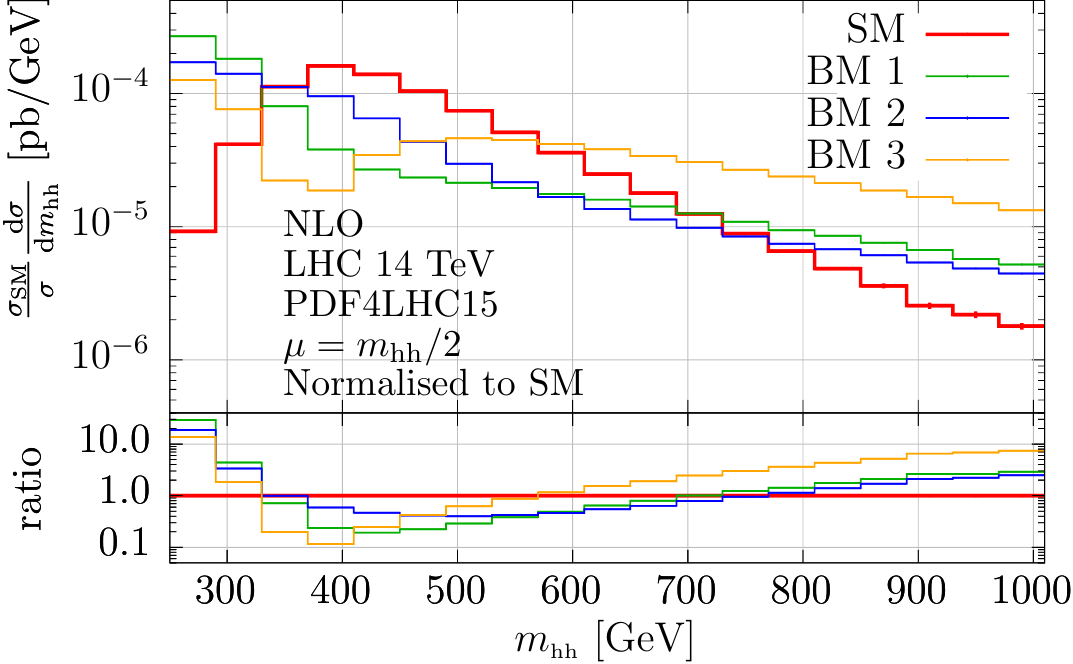}
    \caption{\label{fig:mHH_BM123}}
\end{subfigure}
\begin{subfigure}{0.49\textwidth}
\centering
\includegraphics[width=\textwidth]{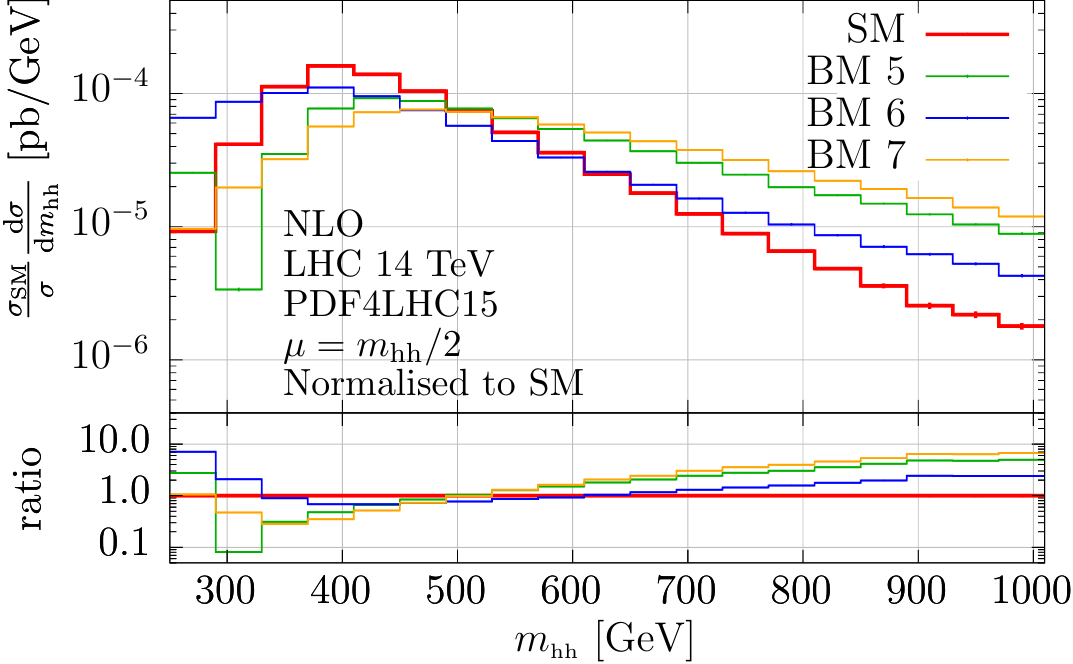}
\caption{  \label{fig:mHH_BM567}}
    \end{subfigure}
    \caption{Normalised Higgs boson pair invariant mass distributions, (a) for benchmark points 1, 2 and 3 compared to the SM, 
    (b) for benchmark points 5, 6 and 7 compared to the SM. All curves are at full
    NLO. The uncertainties shown are statistical only.}
\end{figure}
In Fig.~\ref{fig:mHH_BM123} we show the Higgs boson pair invariant mass distributions  for benchmark
    points 1, 2 and 3 compared to the SM case. The magnitudes of the cross sections are
    similar, due to the fact that the benchmark points were defined
    with the constraint that the total cross section should not exceed
    $6.9\times \sigma_{\rm{SM}}$ at
    13\,TeV~\cite{Capozi:2019xsi}. Nonetheless, the curves are 
    normalised to the SM cross section, such that only shape differences appear in the figure.
    Benchmark point 3 has a value of $\chhh$ where the destructive interference
    between box- and triangle-type contributions is large, which
    leads to the dip in the $\mhh$ spectrum, while the tail is
    enhanced due to non-zero $\cg$ and $\cgg$ values.
    Benchmark point 1 shows the largest enhancement of the very low $\mhh$ region, even though its value for
    $\chhh$ is smaller than the one for benchmark point 2.
    This behaviour can be attributed to the interplay with the nonzero value of $\ctt$, as can be concluded from the analysis in Ref.~\cite{Capozi:2019xsi}.

In Fig.~\ref{fig:mHH_BM567} the $\mhh$ distribution for benchmark
points 5, 6 and 7 is shown, normalised to the SM cross section. 
 Benchmark point 5 shows a narrow dip below
 $\mhh=2m_t$, which would not be present for $\chhh=3.95$ if all other
 couplings were SM-like. In fact, from the analysis in
 Ref.~\cite{Capozi:2019xsi} it can be inferred that the negative
 $\cgg$ value in combination with $\chhh=3.95$ is causing this
 dip in the shape. 

\begin{figure}[htb]
  \centering
  \begin{subfigure}[b]{0.49\textwidth}
    \includegraphics[width=\textwidth]{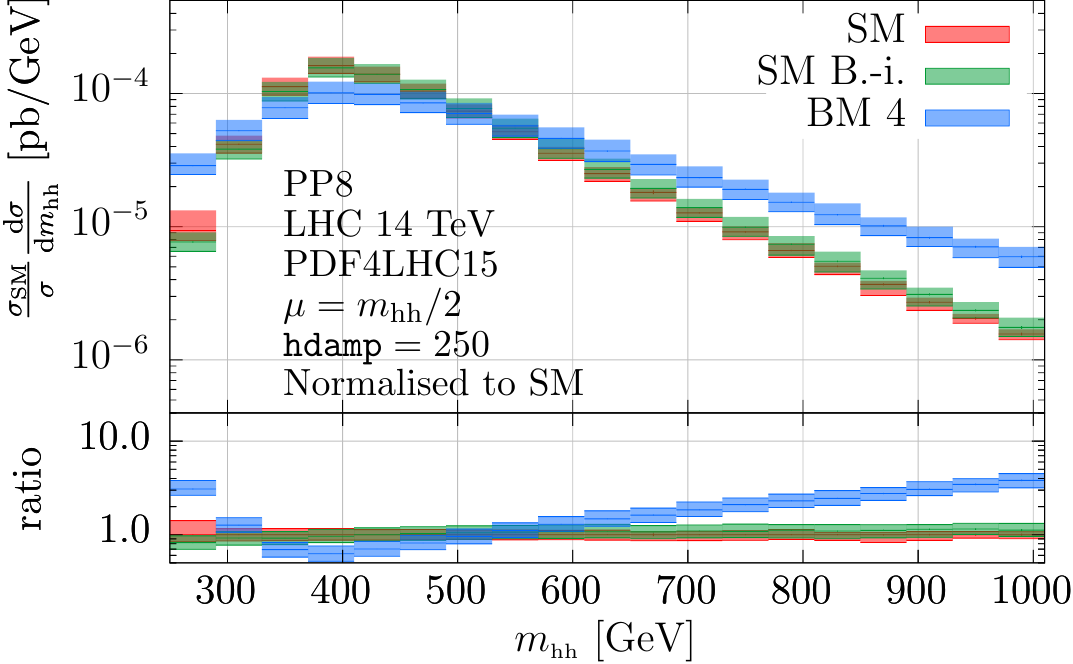}
    \caption{\label{fig:mHH_BM4}}
\end{subfigure}
\begin{subfigure}[b]{0.49\textwidth}
\centering
    \includegraphics[width=\textwidth]{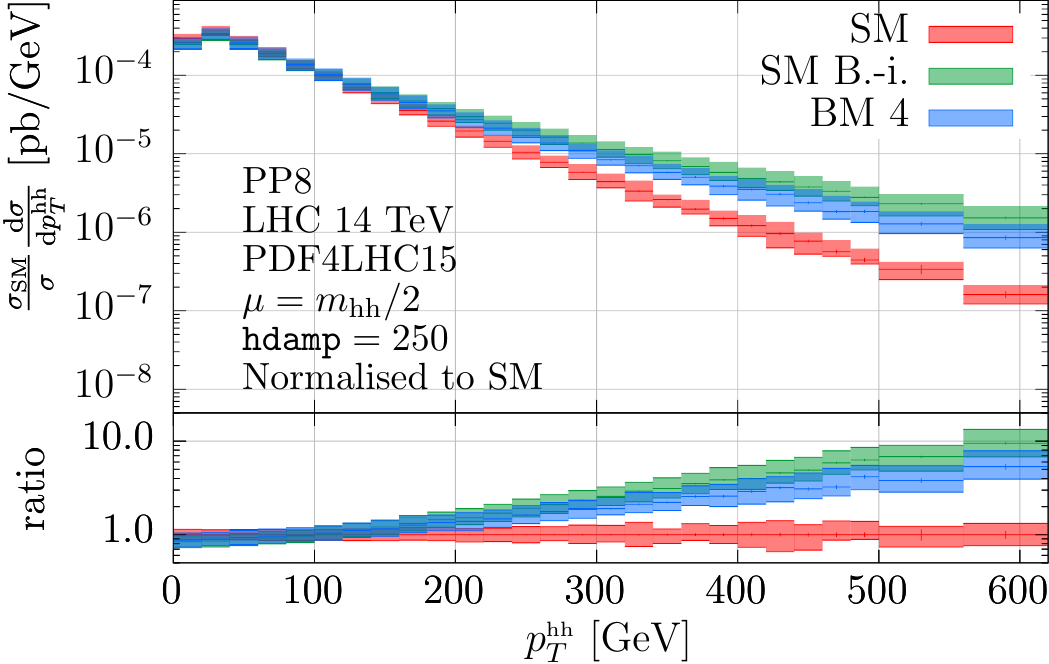}
    \caption{\label{fig:pTHH_BM4}}
\end{subfigure}
    \caption{Normalised distributions for the SM
    compared to benchmark point 4, both at NLO matched to \pythia{} (PP8). For the SM case, results with
    the full $m_t$-dependence and in the Born-improved (B.-i.) $m_t\to\infty$ approximation are shown.
    (a)  Higgs boson pair invariant mass distribution, (b) transverse momentum of the Higgs boson pair.}
\end{figure}
In Fig.~\ref{fig:mHH_BM4}  we consider  benchmark point 4 compared to
the full NLO SM as well as in the Born-improved $m_t\to\infty$ limit,
matched to \pythia{} in all cases. The curves for the Born-improved HTL SM case and
for benchmark point 4 are normalised to the SM cross section. Even though the  $m_t\to\infty$
approximation shows an enhanced tail compared to the full SM, the
enhancement of the tail in the case of benchmark 4 is much more
pronounced. The situation is different for the $\pthh$ distribution,
shown in Fig.~\ref{fig:pTHH_BM4}. For this observable, the
results for  benchmark 4 and the Born-improved $m_t\to\infty$
approximation are very close. This fact again shows the importance of
the full NLO corrections in order to clearly identify new physics
effects. 

In both Fig.~\ref{fig:mHH_BM4} and Fig.~\ref{fig:pTHH_BM4}, in order to obtain the scale uncertainty bands, the variation curves 
were normalised by the ratio of the central-scale prediction to the SM cross section. Thus the bands have
the same relative size as in an unnormalised plot.
We also investigated a different option to produce the scale bands for the normalised cross section, where the scale uncertainties are not normalised by the ratio $\sigma_{\rm{SM}}/\sigma(\mu_0)$, but rather by $\sigma_{\rm{SM}}/\sigma(c\,\mu_0),\;c \in \{0.5,1,2\}$, i.e. by their own cross section at the considered scale.
For the $\mhh$ distribution this type of normalisation makes the scale bands disappear within the statistical uncertainties. This is because for central scale choice $\mu_0=\mhh/2$ the scale variations do not introduce significant shape changes for this observable. 
For the $\pthh$ distribution the situation is different, firstly because the central scale choice is not aligned with the observable, and secondly because the tail of the $\pthh$ distribution is dominated by $HH$+jet events, which are the leading order in this channel and therefore show larger scale uncertainties.
Indeed we observe that when normalised to their own cross-section prediction, the scale uncertainty bands differ from the central prediction by $\pm(3\mbox{-}4)\%$ for $\pthh \lesssim 200$\,GeV, and up to $\mp 18\%$ at $\pthh =600$\,GeV. This is to be compared to an overall scale uncertainty of $30\mbox{-}40 \%$ at $\pthh =600$\,GeV with the default normalisation method.
To confirm this interpretation, we also investigated scale-varied differential cross sections for the distribution of the transverse momentum of one (any) of the Higgs bosons $p_T^h$, which is an observable that is not aligned with the choice of scale $\mu=\mhh/2$, but which does get contributions from genuine radiative corrections at NLO. When normalised to their own cross section, the scale-varied predictions differ by $\pm 1\%$ from the central prediction at low-to-moderate $p_T^h$, and grow up to $\mp(8\mbox{-}10)\%$ in the tail at $p_T^h = 600$\, GeV. As expected, the scale uncertainties are non-vanishing but still smaller than for the $\pthh$ distribution. In comparison, the full (unnormalised) scale uncertainties are of the order of $\pm (20\mbox{-}25)\%$ across the $p_T^h$ range.

\begin{figure}[htb]
\centering
   \begin{subfigure}[b]{0.49\textwidth}
     \includegraphics[width=\textwidth]{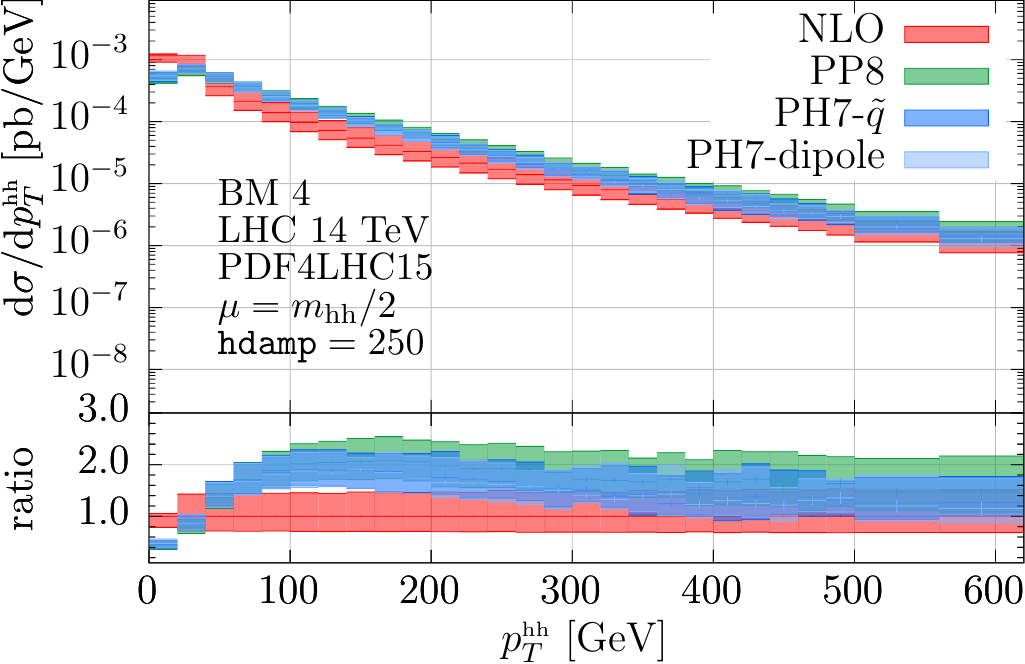}
     \caption{}
     \end{subfigure}
    \begin{subfigure}[b]{0.49\textwidth}
      \includegraphics[width=\textwidth]{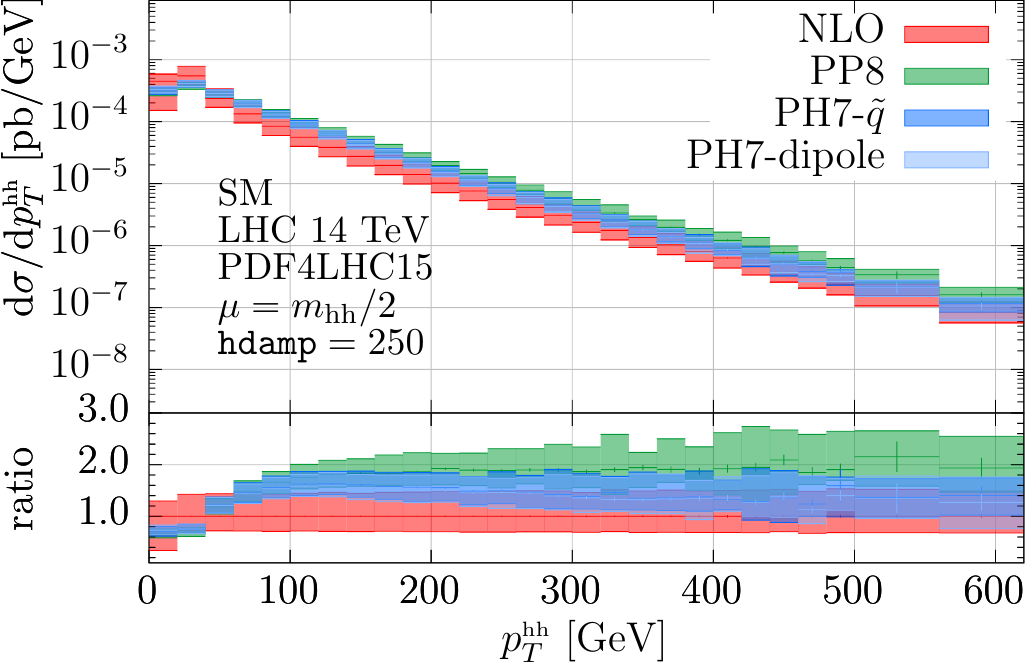}
      \caption{}
     \end{subfigure}
   \caption{(a) Transverse momentum of the Higgs boson pair for benchmark
    point 4,  at NLO matched to \pythia{} (PP8) and two
    different \herwig{} parton showers (PH7-$\tilde{q}$ and PH7-dipole), compared to the fixed-order
    result (NLO); (b) same as in (a) but for the SM case.
    }
    \label{fig:pTHH_BM4_PS}
\end{figure}
In Fig.~\ref{fig:pTHH_BM4_PS} we compare NLO predictions matched to different 
parton showers, namely \pythia{} and two \herwig{} parton showers 
(the angular-ordered $\tilde{q}$ and the dipole shower), to the fixed-order case, (a)
for benchmark point 4, and (b) for the SM case. We observe that the
enhancement of the tail with \POWHEG+\pythia{} present in the SM case
is much less pronounced for benchmark 4, where the \POWHEG+\pythia{} result
also touches onto the fixed-order result at large $\pthh$.
This behaviour is most likely due to the non-zero values for $\cg$ and
$\cgg$ for benchmark 4, which cause the tail of the distribution
already to be harder than in the SM, such that additional hard
radiation created by the shower has a lower relative impact.

\clearpage

\section{Conclusions}
\label{sec:conclusions}

We have presented a publicly available implementation of Higgs boson pair production in gluon
fusion within an Effective Field Theory framework, calculated at full NLO QCD, in the \POWHEGBOXV{}.
The code allows five anomalous couplings relevant for di-Higgs
production to be varied and offers the possibility to produce fully differential final states.

We have also investigated the behaviour of the shape of the invariant
mass distribution of the Higgs boson pair, $\mhh$, for seven benchmark points
representing characteristic $\mhh$-shapes based on an NLO analysis~\cite{Capozi:2019xsi}.
In addition, for one of the benchmark points (benchmark 4), characterised by an enhanced tail in the $\mhh$ distribution,
we carried out a comparison to the full NLO SM result as well as to the $m_t\to \infty$ approximation, including scale uncertainties, for both the $\mhh$ and the $\pthh$ distributions.
We found that the two distributions show different characteristics concerning the distinction of the BSM curve from the SM (and approximate SM) curves: 
while in the  $\mhh$ distribution the enhanced tail of benchmark 4 is clearly outside the uncertainty bands of the SM predictions, in the $\pthh$ distribution the bands for benchmark 4 and the SM in the $m_t\to \infty$ approximation overlap. This again demonstrates the importance of using full NLO predictions, particularly in trying to resolve partly degenerate directions in the space of anomalous couplings.

We further produced results for the  $\pthh$ distribution  matched to three different parton showers: \pythia{} and two different \herwig{} parton showers. The two \herwig{} parton showers show very similar results.
From previous SM results, it is known that loop-induced processes like $gg\to HH$ in \POWHEG{} matched to \pythia{} can show a harder tail than with other parton showers.
However, in the case of benchmark point 4,  \pythia{} produces less additional hard radiation than in the SM case, such that the \pythia{}  and \herwig{} results are much more similar.

Our studies show that the behaviour of higher-order effects known from the SM does not necessarily carry over to the BSM
case, such that precise predictions for both cases are necessary to clearly identify new physics effects.
With our code, fully exclusive studies of anomalous couplings in Higgs boson pair production at NLO QCD are possible.

\section*{Acknowledgements}
We would like to thank Gerhard Buchalla for useful discussions and
Matteo Capozi and Gionata Luisoni for collaboration on earlier
versions of this code. We also are grateful to Tom\'a\v{s} Je\v{z}o
and Emanuele Re for very useful comments about the \POWHEGBOX.
This research was supported in part by the COST Action CA16201 (`Particleface') of the European Union
and  by  the  Deutsche  Forschungsgemeinschaft (DFG, German Research
Foundation) under grant 396021762 - TRR 257. LS is supported by the Royal Society under 
grant number RP\textbackslash R1\textbackslash 180112.
MK acknowledges support by the Swiss National  Science  Foundation  (SNF)  under  grant  number  200020-175595
and by the Forschungskredit of the University of Zurich, grant no. FK-19-102.
We also acknow\-ledge
resources provided by the Max Planck Computing and Data
Facility (MPCDF).


\renewcommand \thesection{\Alph{section}}
\renewcommand{\theequation}{\Alph{section}.\arabic{equation}}
\setcounter{section}{0}
\setcounter{equation}{0}

\section{Appendix}

\subsection{Running with full top quark mass dependence ({\tt mtdep=3})}
\label{sec:appendix1}

In this appendix we give some further details about the running mode with full top quark mass dependence.

The two-loop virtual amplitudes in the NLO calculation with full top quark mass dependence are computed via a grid
which encodes the dependence of the virtual two-loop amplitude on the
kinematic invariants $\hat{s}$ and
$\hat{t}$~\cite{Heinrich:2017kxx}. We emphasize that the numerical
values $m_H=125$\,GeV and $m_t=173$\,GeV are hardcoded in this
grid and therefore should not be changed in {\tt powheg.input-save} when running in the {\tt mtdep=3} mode.
The grid is generated using python code and is directly interfaced to the
\POWHEGBOX{} fortran code via a python/C API. In order for the grid to be found by the code, the
files ({\tt events.cdf, createdgrid.py, Virt\_full\_*E*.grid}) from the
folder {\tt Virtual} need to be copied into the local folder where the
code is run. Instead of copying the files, we suggest to create a
symbolic link to the needed files.
All this is done automatically if you use the script  {\tt run.sh}.

To do this manually:
assuming the code is run from a
subfolder (e.g. {\tt testrun}) of the process folder, the link can be created in this subfolder as follows:
\begin{description}
\item{\tt ln -s ../Virtual/events.cdf events.cdf}
\item{\tt ln -s ../Virtual/creategrid.py creategrid.py}
\item{\tt for grid in ../Virtual/Virt\_full\_*E*.grid; do ln -s \$grid; done}
\end{description}
Once the links are in place, the code can be run with {\tt mtdep=3} as usual.  
The python code {\tt creategrid.py} will then combine the virtual
grids generated with the 23 combinations of coupling values
to produce a new file {\tt Virt\_full\_*E*.grid} corresponding to the
values of $\chhh, \ct, \ctt, \cg, \cgg$ defined by the user in the {\tt powheg.input-save} file.

The python code for the grid relies on the {\tt numpy} and {\tt sympy} packages, which 
the user should install separately. When building the {\tt ggHH} process the Makefile will find the embedded python 3 library
via a call to {\tt python3-config}, which the user should ensure is configured correctly and points to the correct library.
Note that on some systems the python/C API does not search for packages (such as {\tt numpy} and {\tt sympy}) in the same 
paths as the python executable would, the user should ensure that these packages can be found also by an embedded python program.
To ensure that the linked files are found, we recommend to add the run subfolder to {\tt PYTHONPATH}.

\subsection{Powheg input and run scripts}

The {\tt run.sh} script in the {\tt testrun} folder allows 
the different stages of \POWHEG{} to be run easily.
By typing {\tt ./run.sh} without any argument a menu with the
4 {\tt mtdep} running modes described above is shown. 
For all {\tt mtdep} running modes, {\tt run.sh} will make the code go through 
the various steps (parallel stages) of the calculation: 
\begin{description}
 \item[{\tt parallelstage=1}:]
generation of the importance sampling grid for the Monte Carlo integration; 
 \item[{\tt parallelstage=2}:] calculation of the integral for the inclusive cross section and an upper bounding function of the integrand;
 \item[{\tt parallelstage=3}:] upper bounding factors for the generation of radiation are computed;
 \item[{\tt parallelstage=4}:] event generation, i.e. production of {\tt pwgevents-*.lhe} files.
\end{description}

\noindent Please note: if you use the script {\tt run.sh [mtdep]}, the value for {\tt mtdep} given as an argument to {\tt run.sh} will be used, even if you specified a different value for {\tt mtdep} in {\tt powheg.input-save}.

After running {\tt parallelstage=4}, the LHE files produced by \POWHEG{} can be directly showered by either \pythia{} or \herwig{}. We provide a minimal setup for producing parton-shower matched distributions in {\tt test-pythia8}, respectively {\tt test-herwig7}. Both the angular-ordered and the dipole-shower implemented in \herwig{} can be used by changing the {\tt showeralg} flag to either {\tt default} or {\tt dipole} in {\tt HerwigRun.sh}.

Further, we should point out that 
\POWHEG{} offers the possibility to use a damping factor $h=\texttt{hdamp}$ of the
form~\cite{Alioli:2008tz,Alioli:2009je}
\begin{align}
  F=\frac{h^{2}}{(\pthh)^2+h^{2}}\,,
\end{align}
where $\pthh$ is the transverse momentum of the Higgs boson pair, to
limit the amount of hard radiation which is exponentiated in the
Sudakov form factor. The setting $F\equiv1$, corresponding to {\tt hdamp}$=\infty$, results in
quite hard tails for observables like
$\pthh$~\cite{Heinrich:2017kxx,Heinrich:2019bkc}. Changing the damping factor $F$ by
setting the flag {\tt hdamp} to some finite value in the input
card softens the high transverse momentum tails. Varying {\tt hdamp} allows shower uncertainties to be assessed
within the \POWHEG{} matching scheme. However, 
{\tt hdamp} should not be so low that it starts to cut into the Sudakov
regime. In fact, a too low value for {\tt hdamp} could spoil the
logarithmic accuracy of the prediction. For this reason we suggest not
to choose values for {\tt hdamp} below $\sim 200$. Our default value is  {\tt hdamp=250}.

\bibliographystyle{JHEP}

\bibliography{refs_full_EWChL}

\end{document}